\documentclass[amsmath,amssymb,aps]{revtex4}
\usepackage{txfonts}
\usepackage{epsfig}

\newcommand{\be}{\begin{equation}}
\newcommand{\ee}{\end{equation}}
\newcommand{\bel}[1]{\begin{equation}\label{#1}}
\newcommand{\bea}{\begin{eqnarray}}
\newcommand{\eea}{\end{eqnarray}}
\newcommand{\ba}{\begin{array}}
\newcommand{\ea}{\end{array}}

\begin{document}

\title{Current Distribution and random matrix ensembles for an integrable
asymmetric fragmentation process}

\author{A.\ R\'akos and G.M. Sch\"utz}

\affiliation{Institut f\"ur Festk\"orperforschung, Forschungszentrum 
J\"ulich - 52425 J\"ulich, Germany}

\date{\today}

\begin{abstract}
  We calculate the time-evolution of a discrete-time fragmentation
  process in which clusters of particles break up and reassemble and
  move stochastically with size-dependent rates. In the
  continuous-time limit the process turns into the totally asymmetric
  simple exclusion process (only pieces of size 1 break off a given
  cluster).  We express the exact solution of master equation for the
  process in terms of a determinant which can be derived using the
  Bethe ansatz.  From this determinant we compute the distribution of
  the current across an arbitrary bond which after appropriate scaling
  is given by the distribution of the largest eigenvalue of the
  Gaussian unitary ensemble of random matrices.  This result confirms
  universality of the scaling form of the current distribution in the
  KPZ universality class and suggests that there is a link between integrable
  particle systems and random matrix ensembles.
\end{abstract}

\maketitle

\section{Introduction}
Asymmetric exclusion processes are paradigmatic models for systems far
from equilibrium, both for their wide range of applications and the
availability of exact results \cite{Ligg99,Schu00}.  Despite their
greatly reduced complexity they capture various generic nonequilibrium
phenomena such as the occurrence of shocks and particle condensation
\cite{Evan00,Muka00,Schu03}. A major breakthrough in the exact
calculation of universal properties came with the realization that
various ensembles of random matrices occur in the study of current
fluctuations and related quantities in the totally asymmetric simple
exclusion process (TASEP) \cite{Joha00,Prae02a} and also in the polynuclear
growth model \cite{Prae00,Baik00,Prae02b,Sasa03}. Here we use random matrix
theory combined with the Bethe ansatz to study a process introduced by
Priezzhev \cite{Prie02} which describes the stochastic fragmentation
and reassembly of diffusing particle clusters in discrete time.

In the totally asymmetric fragmentation process \cite{Prie02} one considers
a one-dimensional lattice where each lattice point is occupied by at
most one particle. A string of $n$ consecutive particles
(and bounded by empty sites) is considered a cluster of size $n$. 
The stochastic time evolution occurs in discrete time steps. From
each cluster a piece of size $n'<n$ may break off and move to the
right by one lattice unit with probability $(1-p)p^{n'}$. The whole cluster
may hop with probability $p^n$. According to this definition 
the transition rules are the following:
\begin{align*}
0\underbrace{AA \dots AAA}_{n}0 \rightarrow 0\underbrace{AA\dots AA}_{n-1}0A & 
\quad \text{ with probability } p(1-p) \cr
0\underbrace{AA \dots AAA}_{n}0 \rightarrow 0\underbrace{AA\dots A}_{n-2}0AA & 
\quad \text{ with probability } p^2(1-p) \cr
\vdots & \cr
0\underbrace{AA \dots AAA}_{n}0 \rightarrow 0A0\underbrace{A\dots AAA}_{n-1} & 
\quad \text{ with probability } p^{n-1}(1-p) \cr
0\underbrace{AA \dots AAA}_{n}0 \rightarrow 00\underbrace{AA\dots AAA}_{n} & 
\quad \text{ with probability } p^n 
\end{align*} 
Notice that two clusters can merge through hopping if they are
separated by only one vacant site. Other fragmentation processes have
been studied recently \cite{Khor99,Maju00,Scha02,Barm02}.

In the noiseless limit $p=1$ neither fragmentation nor
recombination occurs and all clusters
move ballistically with probability 1. In the limit $p\to 0$ with an
appropriate rescaling of time ($t \to \infty$ with $t' = pt$ fixed)
only pieces of size 1, i.e., single particles may break off. This process
is equivalent to the usual TASEP.
In discrete time the fragmentation process may be interpreted as an
asymmetric exclusion process with long-range hopping. This is
forbidden in the usual discrete-time TASEP with parallel update
studied in Ref. \cite{Joha00}. Moreover, the fragmentation process
has no particle-hole symmetry and the stationary distribution is a
product measure with constant density $\rho$ and stationary current
\bel{1}
j = p \frac{\rho(1-\rho)}{1-p\rho}.
\ee
The absence of correlations implies that fragmentation and recombination
balance each other such that no condensation into macroscopic clusters
occurs. Indeed, our interest is not in the stationary state of the system
but how it evolves from a fully "phase-separated" initial state where the
left half of the infinite lattice (all sites  $k \leq 0$) is occupied
while the right half (all sites  $k > 0$) is vacant.

Under Eulerian scaling (lattice constant $a$ and time step $\tau$
tend to zero with the ratio $a/\tau$ kept constant) one expects
the coarse-grained density $\rho(x,t)$ to be governed by the hydrodynamic
conservation law
\bel{2}
\partial_t \rho + \partial_x j(\rho) =0
\ee
with the macroscopic current (\ref{1}). Our interest is in the microscopic
fluctuations of the current which after appropriate rescaling is
expected to be given by an universal scaling function. In particular, one 
obtains universal corrections to (\ref{2}) below the Euler scale.

In Sec. 2 we first consider the continuous-time limit of the model,
i.e., the TASEP.  We present a new derivation of the result Eq. (1.18)
of Ref. \cite{Joha00} for the exact distribution of the
time-integrated current.  We do not use the combinatorial arguments
employed by Johansson, but show how the same expression can be
obtained logically independently from the Bethe ansatz solution via
the determinant expression of Ref. \cite{Schu97}.  In Sec. 3 we extend
this approach to calculate the current distribution for the
fragmentation process with arbitrary fragmentation parameter $p$.  In
Sec. 4 we present some conclusions. Some properties of the functions
we use in the calculation are given in the appendices.

\section{Continuous time TASEP}
\subsection{Known results}
\subsubsection{Current fluctuations}  
\label{RM}
Consider the TASEP on an infinite chain in continuous time
where particles hop to the right with rate 1, provided the target site is
vacant. At
$t=0$ the left half of the system (from site $-\infty$ to site $0$) is
occupied while the right half is empty. We focus on the probability $P(M,N,t)$
that the $N$th particle (which was on the $(1-N)$th site of the infinite
cluster at $t=0$)
hops at least $M$ times up to time $t$. Using combinatorial arguments
involving the longest increasing subsequences of random
permutations Johansson has proved (Eq. (1.18) in Ref. \cite{Joha00})
\begin{equation}
\label{P}
P(M,N,t)=\frac{1}{Z'_{M,N}}\int_{[0,t]^N} \prod_{1\leq i<j\leq N} (x_i-x_j)^2
\prod_{j=1}^N x_j^{M-N}e^{-x_j}d^Nx 
\end{equation} 
for $M\geq N$. To have proper normalization the partition sum has
to be
\begin{equation}
Z'_{M,N}=\int_{[0,\infty]^N} \prod_{1\leq i<j\leq N} (x_i-x_j)^2
\prod_{j=1}^N x_j^{M-N}e^{-x_j}d^Nx.
\end{equation}
The expression (\ref{P}) is equal to the probability that the largest
eigenvalue of a random matrix $AA^\ast$ is $\leq t$ where $A$ is
a $N \times M$ matrix of complex Gaussian random variables with mean zero
and variance 1/2 \cite{Jame64}.

Let $J(x,t)$ be the number of particles that have crossed the lattice bond
$(x,x+1)$ up to time $t$, i.e., the time-integrated current. By construction
one has for the probability that $J(x,t) > m$
\bel{3}
\text{Prob}\left[J(x,t) > m\right] = P(m+x+1,m+1,t).
\ee

\subsubsection{Bethe ansatz}
\label{BA}

Consider the TASEP on the infinite chain with a finite ($N$) number of particles
located initially at sites $A_N=\{l_1,l_2, \cdots ,l_N\}$ ($l_1<l_2< \dots
<l_N$). It is proved in 
\cite{Schu97} that the probability of having these particles on sites
$B_N=\{k_1,k_2, \cdots ,k_N\}$ ($k_1<k_2< \cdots <k_N$) at time $t$ is
given by the determinant
\begin{equation}
\label{Q}
Q(A_N,B_N;t)=
\left| 
\begin{array}{cccc}
F_0(k_1-l_1;t) & F_{-1}(k_1-l_2;t) & \cdots &  F_{-N+1}(k_1-l_N;t) \\
F_1(k_2-l_1;t) & F_0(k_2-l_2;t)    & \cdots & F_{-N+2}(k_2-l_N;t) \\
\vdots         & \vdots            &        & \vdots \\
F_{N-1}(k_N-l_1;t) & F_{N-2}(k_N-l_2;t) & \cdots & F_0(k_N-l_N;t)
\end{array}
\right|.
\end{equation}
For the definition and some properties of the $F_p(n;t)$ functions see
appendix \ref{F_functions}. 

The determinant has been obtained from a coordinate Bethe ansatz for the 
conditional probability $Q(A_N,B_N;t)$. Given the determinant, 
the proof that it is the solution of the master equation
for the TASEP follows from standard relations for determinants, 
for details see \cite{Schu97}. In the next subsection we show how
(\ref{P}) can be derived directly from (\ref{Q}) without reference to
combinatorial properties of the process.

\subsection{Calculation}
\label{Calculation}

The dynamics of the rightmost $N$ particles in the TASEP are
independent of all particles to their left. Therefore $P(M,N,t)$ of 
sec.~\ref{RM} can be expressed via $Q(A_N,B_N;t)$ of sec.~\ref{BA}:
\begin{equation}
\label{PMNt}
P(M,N,t)= \\
\sum_{M-N<k_1<k_2< \cdots <k_N} Q(\{-N+1,-N+2, \cdots ,0\},
\{k_1,k_2, \cdots ,k_N\};t).
\end{equation}
Inserting (\ref{P}) and (\ref{Q}) one gets
\begin{multline}
\frac{1}{Z'_{M,N}}\int_{[0,t]^N} \prod_{1\leq i<j\leq N} (x_i-x_j)^2
\prod_{j=1}^N x_j^{M-N}e^{-x_j}d^Nx = \\
\sum_{M-N<k_1<k_2< \cdots <k_N} 
\left| 
\begin{array}{cccc}
F_0(k_1+N-1;t) & F_{-1}(k_1+N-2;t) & \cdots &  F_{-N+1}(k_1;t) \\
F_1(k_2+N-1;t) & F_0(k_2+N-2;t)    & \cdots & F_{-N+2}(k_2;t) \\
\vdots         & \vdots            &        & \vdots \\
F_{N-1}(k_N+N-1;t) & F_{N-2}(k_N+N-2;t) & \cdots & F_0(k_N;t)
\end{array}
\right|.
\label{to_show}
\end{multline}
In what follows we show by determinant manipulations and
using properties of the functions $F_k$ that this equality holds.

Our starting point is the rhs.\ of (\ref{to_show}). The summation over
$(k_1,k_2, \cdots k_N)$ can be done in $N$ steps for which we choose
the following sequence:
\begin{equation}
\sum_{M-N<k_1<k_2< \cdots <k_N} = \sum_{k_N=M}^\infty
\sum_{k_{N-1}=M-1}^{k_N-1} \cdots \sum_{k_2=M-N+2}^{k_3-1} 
\sum_{k_1=M-N+1}^{k_2-1}
\end{equation}
After summation over $k_1$ (for which we use (\ref{sum1})) the first
row of the matrix becomes 
\begin{equation}
    F_1(M;t)-F_1(k_2+N-1;t) \quad F_0(M-1;t)-F_0(k_2+N-2;t) \quad
 \cdots \quad F_{-N+2}(M-N+1;t)-F_{-N+2}(k_2;t)
\end{equation} 
which reduces to
\begin{equation}
F_1(M;t) \quad F_0(M-1;t) \quad \cdots \quad F_{-N+2}(M-N+1;t)
\end{equation} 
after adding the second row to it. The same method can be used up to
the sum over $k_{N-1}$. For the last sum we use (\ref{sum3}) and
finally we get 
\begin{equation}
\label{after_sum}
\left|
\begin{array}{cccc}
F_1(M;t) & F_0(M-1;t) & \cdots & F_{-N+2}(M-N+1;t) \\
F_2(M+1;t) & F_1(M;t) & \cdots & F_{-N+3}(M-N+2;t) \\
\vdots & \vdots & & \vdots  \\
F_N(M+N-1;t) & F_{N-1}(M+N-2;t) & \cdots & F_1(M;t)
\end{array}
\right|
\end{equation} 
for the rhs.\ of (\ref{to_show}). 

It turns out to be useful to represent all the $F$ functions in
(\ref{after_sum}) by an integral using (\ref{int}). Since for
$M\geq N$ all 
are zero at $t=0$ the determinant
(\ref{after_sum}) can be written as  
\begin{equation}
\left|
\begin{array}{cccc}
\int_0^t F_0(M-1;\tau)d\tau & \int_0^t F_{-1}(M-2;\tau)d\tau & \cdots
& \int_0^t F_{-N+1}(M-N;\tau)d\tau \\ 
\int_0^t F_1(M;\tau)d\tau & \int_0^t F_0(M-1;\tau)d\tau & \cdots &
\int_0^t F_{-N+2}(M-N+1;\tau)d\tau \\ 
\vdots & \vdots & & \vdots  \\
\int_0^t F_{N-1}(M+N-2;\tau) d\tau & \int_0^t F_{N-2}(M+N-3;\tau)d\tau
& \cdots & \int_0^t F_0(M-1;\tau)d\tau 
\end{array}
\right|
\end{equation} 
In the second row we perform a partial integration after which the $i$th 
element becomes
\begin{equation}
tF_{2-i}(M-i+1;t)-\int_0^t \tau F_{1-i}(M-i;\tau)d\tau 
\end{equation} 
Note that the constant part is $t$ times the corresponding element of
the first row so after substracting $t$ times the first row the second
row becomes 
\begin{equation}
-\int_0^t \tau F_0(M-1;\tau)d\tau \quad -\int_0^t \tau
 F_{-1}(M-2;\tau)d\tau \quad \cdots \quad -\int_0^t \tau
 F_{-N+1}(M-N;\tau)d\tau.  
\end{equation}
In the third row we perform a double partial integration then we add $t^2/2$ 
times the first row and substract $t$ times the (original) second row. 
Repeating the same procedure for all of the rows we get
\begin{multline}
\left|
\begin{array}{cccc}
\int_0^t F_0(M-1;\tau)d\tau & \int_0^t F_{-1}(M-2;\tau)d\tau & \cdots
& \int_0^t F_{-N+1}(M-N;\tau)d\tau \\ 
-\int_0^t \tau F_0(M-1;\tau)d\tau & -\int_0^t \tau
F_{-1}(M-2;\tau)d\tau & \cdots & -\int_0^t \tau
F_{-N+1}(M-N;\tau)d\tau \\ 
\frac{1}{2}\int_0^t \tau^2 F_0(M-1;\tau)d\tau & \frac{1}{2}\int_0^t
\tau^2 F_{-1}(M-2;\tau)d\tau & \cdots & \frac{1}{2}\int_0^t \tau^2
F_{-N+1}(M-N;\tau)d\tau \\ 
-\frac{1}{3!}\int_0^t \tau^3 F_0(M-1;\tau)d\tau &
-\frac{1}{3!}\int_0^t \tau^3 F_{-1}(M-2;\tau)d\tau & \cdots &
-\frac{1}{3!}\int_0^t \tau^3 F_{-N+1}(M-N;\tau)d\tau \\ 
\vdots & \vdots & & \vdots  \\
\frac{\pm 1}{(N-1)!}\int_0^t \tau^{N-1} F_0(M-1;\tau)d\tau &
\frac{\pm 1}{(N-1)!}\int_0^t \tau^{N-1} F_{-1}(M-2;\tau)d\tau & \cdots
& \frac{\pm 1}{(N-1)!}\int_0^t \tau^{N-1} F_{-N+1}(M-N;\tau)d\tau  
\end{array}
\right|= \\ \\
(-1)^{\left[\frac{N}{2}\right]}\prod_{i=1}^{N-1}\frac{1}{i!}\int_{[0,t]^N} 
\tau_1^0 \tau_2^1 \tau_3^2 \tau_4^3 \cdots \tau_N^{N-1} 
\left|
\begin{array}{cccc}
F_0(M-1;\tau_1) & F_{-1}(M-2;\tau_1) & \cdots & F_{-N+1}(M-N;\tau_1) \\
 F_0(M-1;\tau_2) & F_{-1}(M-2;\tau_2) & \cdots &  F_{-N+1}(M-N;\tau_2) \\
 F_0(M-1;\tau_3) &  F_{-1}(M-2;\tau_3) & \cdots & F_{-N+1}(M-N;\tau_3) \\
\vdots & \vdots & & \vdots  \\
 F_0(M-1;\tau_N) &  F_{-1}(M-2;\tau_N) & \cdots & F_{-N+1}(M-N;\tau_N)
\end{array}
\right|d^N\tau.
\label{res}
\end{multline} 

Using (\ref{sum2}) we can rewrite $F_{-1}(M-2;\tau_i)$ as
$F_0(M-2;\tau_i)-F_0(M-1;\tau_i)$.  So adding the first
column to the second one the $i$th element of the latter  becomes
$F_0(M-2;\tau_i)$. Similar transformations can be done with the other
columns as well: since according to (\ref{pneg})
\begin{equation}
F_p(n;t) =\sum_{m=0}^{-p} (-1)^m \binom{-p}{m} F_0(n+m;t) \quad \text{for } 
p\leq 0,
\end{equation} 
one can
add suitable linear combinations of the first $l-1$ columns to the $l$th 
one so that the $i$th element of the $l$th column becomes 
\begin{equation}
F_0(M-l;\tau_i)=e^{-\tau_i} \frac{\tau_i^{M-l}}{(M-l)!}.
\end{equation} 
After these transformations one gets for (\ref{res}):
\begin{equation}
(-1)^{\left[\frac{N}{2}\right]}\prod_{i=1}^{N-1}\frac{1}{i!}
\prod_{i=1}^{N}\frac{1}{(M-i)!}
\int_{[0,t]^N} \prod_{i=1}^N \left( \tau_i^{M-N}e^{-\tau_i}\right)  
\tau_1^0 \tau_2^1 \tau_3^2 \tau_4^3 \cdots \tau_N^{N-1} 
\left|
\begin{array}{ccccc}
\tau_1^{N-1} & \tau_1^{N-2} & \cdots & \tau_1^1 & \tau_1^0 \\
\tau_2^{N-1} & \tau_2^{N-2} & \cdots & \tau_2^1 & \tau_2^0 \\
\vdots & \vdots & & \vdots & \vdots \\
\tau_N^{N-1} & \tau_N^{N-2} & \cdots & \tau_N^1 & \tau_N^0 \\
\end{array}
\right| d^N\tau.
\end{equation} 

Since the integration is symmetric in the $\tau_i$ while the determinant
is antisymmetric we may replace the product of the $\tau_i$ by antisymmetric 
combination
\begin{equation}
\frac{1}{N!}\left|
\begin{array}{ccccc}
\tau_1^{0} & \tau_1^{1} & \cdots & \tau_1^{N-2} & \tau_1^{N-1} \\
\tau_2^{0} & \tau_2^{1} & \cdots & \tau_2^{N-2} & \tau_2^{N-1} \\
\vdots & \vdots & & \vdots & \vdots \\
\tau_N^{0} & \tau_N^{1} & \cdots & \tau_N^{N-2} & \tau_N^{N-1} \\
\end{array}
\right| =
\frac{(-1)^{\left[\frac{N}{2}\right]}}{N!}
\left|
\begin{array}{ccccc}
\tau_1^{N-1} & \tau_1^{N-2} & \cdots & \tau_1^1 & \tau_1^0 \\
\tau_2^{N-1} & \tau_2^{N-2} & \cdots & \tau_2^1 & \tau_2^0 \\
\vdots & \vdots & & \vdots & \vdots \\
\tau_N^{N-1} & \tau_N^{N-2} & \cdots & \tau_N^1 & \tau_N^0 \\
\end{array}
\right|.
\end{equation} 
This is the Vandermonde determinant
\begin{equation}
\label{lem}
\left|
\begin{array}{ccccc}
\tau_1^{N-1} & \tau_1^{N-2} & \cdots & \tau_1^1 & \tau_1^0 \\
\tau_2^{N-1} & \tau_2^{N-2} & \cdots & \tau_2^1 & \tau_2^0 \\
\vdots & \vdots & & \vdots & \vdots \\
\tau_N^{N-1} & \tau_N^{N-2} & \cdots & \tau_N^1 & \tau_N^0 \\
\end{array}
\right|=\prod_{1\leq i<j\leq N}(\tau_i-\tau_j),
\end{equation}
and leads to 
\begin{equation}
\label{result} P(M,N,t) =
\prod_{i=1}^N\left( \frac{1}{i!(M-i)!}\right)\int_{[0,t]^N}
\prod_{i=1}^N \left( \tau_i^{M-N}e^{-\tau_i}\right)\prod_{1\leq
  i<j\leq N}(\tau_i-\tau_j)^2 d^N\tau.
\end{equation}
This is in agreement with the lhs.\ of (\ref{to_show}), moreover 
we get the partition function as a ``by-product'':
\begin{equation}
Z'_{M,N}=\prod_{i=1}^N i!(M-i)!
\end{equation} 

\subsection{Asymptotic form of the density profile}

For the TASEP the current (\ref{1}) reduces to $j=\rho(1-\rho)$ and the
density profile of the step-function initial state evolves on Euler
scale according to
\bel{4}
\rho(x,t) = \frac{1}{2}\left( 1-v \right) 
\ee
where $v=x/t$ and $-1 \leq v \leq 1$ \cite{Rost81}. Outside this range
the density keeps it initial value.

To describe the density profile below Euler scale we use the
asymptotic form of the result (\ref{3}) for the current distribution.
It can be shown that that the formula (\ref{result}) can be written as
a Fredholm determinant with the Meixner kernel which in a proper limit
reduces to the Airy kernel (for details see \cite{Joha00}).  This
implies that in this limit the asymptotic form of $P(M,N,t)$ is given
by the Tracy-Widom distribution \cite{Trac94} of the Gaussian unitary ensemble
($F_\text{GUE}$), viz.
\begin{equation}
\lim_{N\to \infty}P([\gamma N],N,\omega(\gamma)N+\sigma(\gamma)N^{1/3}s)=
F_\text{GUE}(s)
\end{equation} 
with
\begin{equation}
\omega(\gamma)=\left(1+\sqrt{\gamma}\right)^2 \quad \text{and} \quad 
\sigma(\gamma)=\gamma^{-1/6}\left(1+\sqrt{\gamma}\right)^{4/3}.
\end{equation} 
This type of scaling is characteristic for one-dimensional surface-growth 
models of the KPZ universality class (\cite{Prae00}).

For our initial state the integrated current $J(x,t)$ gives the number of 
particles being at sites $k>x$ at time $t$. 
It can be shown using theorem 1.6 of \cite{Joha00} that   
\begin{equation}
\label{asj}
\lim_{t\to \infty}\text{Prob}\left[ J([vt],t)\leq \frac{t}{4}(1-v)^2+2^{-4/3}
\left(1-v^2\right)^{2/3}t^{1/3}s\right]=1-F_\text{GUE}(-s)
\end{equation} 
for $0\leq v<1$.
Because of particle-hole symmetry one has a similar expression for
$-1 \leq v \leq 0$.

We can also consider the corresponding surface growth model where $h(x,t)$ is 
defined as
\begin{equation}
h(x,t)= |x| + 2J(x,t)
\end{equation}  
The asymptotic mean shape of $h([vt],t)/t=(1+v^2)/2$ on Euler scale follows 
directly from (\ref{4}). The deviations can be calculated from
(\ref{asj}):
\begin{equation}
\lim_{t\to \infty}\text{Prob}\left[ \frac{t}{2}(1+v^2)-h([vt],t) < 
2^{-1/3}\left(1-v^2\right)^{2/3}t^{1/3}s\right]=F_\text{GUE}(s)
\end{equation} 
for $0<v<1$. This implies that 
\begin{equation}
\label{ash}
\frac{\frac{t}{2}(1+v^2)-\overline{h}([vt],t)}{2^{-1/3}(1-v^2)^{2/3}t^{1/3}} = -c,
\end{equation} 
where $\overline{h}(x,t)$ is the mean height at position $x$ and time $t$, and 
\bel{5}
-c = \int_{-\infty}^\infty ds \, s F_\text{GUE}'(s) = -1.77109
\ee
is the mean of the distribution $F_\text{GUE}$. Since the 
density $\rho(x,t)$ is $\frac{1}{2}(1-\overline{h}(x,t)+\overline{h}(x-1,t))$ 
expression 
(\ref{ash}) allows us to calculate the correction to the density profile 
(\ref{4})
\begin{equation}
\label{asr}
\rho(vt,t) = \frac{1-v}{2}  +c\frac{2^{2/3}}{3}v\left(1-v^2\right)^{-1/3}t^{-2/3}.
\end{equation} 
The KPZ exponent $2/3$ appearing here is universal.
Note that the coefficient $v\left(1-v^2\right)^{-1/3}$ diverges as $v\to\pm 1$. 
In these singular points we expect a $t^{-1/2}$ correction.

\section{The discrete-time fragmentation process}

\subsection{Complete solution of the master-equation by Bethe ansatz}

The solution is very similar to the one of the continuous time TASEP. 
As in Ref. \cite{Schu97} one constructs
the Bethe solution for the master equation for the conditional probabilities
and recasts the expression in terms of a determinant. Then one proves
by elementary matrix manipulations that the analogue of equation (\ref{Q})
is the following  (for details see \cite{Prie02}):
\begin{equation}
\label{QD}
Q(A_N,B_N;t)=
\left| 
\begin{array}{cccc}
D_0(k_1-l_1;t) & D_{-1}(k_1-l_2;t) & \cdots &  D_{-N+1}(k_1-l_N;t) \\
D_1(k_2-l_1;t) & D_0(k_2-l_2;t)    & \cdots & D_{-N+2}(k_2-l_N;t) \\
\vdots         & \vdots            &        & \vdots \\
D_{N-1}(k_N-l_1;t) & D_{N-2}(k_N-l_2;t) & \cdots & D_0(k_N-l_N;t)
\end{array}
\right|.
\end{equation}
For the definition and the main properties of the $D$ functions see
appendix \ref{appendix:D}.

\subsection{Calculation of $P(M,N,t)$}

We follow the strategy of the previous section, but some care needs to
be taken when manipulating sums rather than the time integrals.
Here $P(M,N,t)$ is again defined by (\ref{PMNt}) but now with the $Q$ of
(\ref{QD}). The summation gives the same result as for the continuous
time version since the corresponding properties of the $D$ functions
are the same as those of the $F$ functions:
\begin{equation}
\label{Dafter_sum}
P(M,N,t)=\left|
\begin{array}{cccc}
D_1(M;t) & D_0(M-1;t) & \cdots & D_{-N+2}(M-N+1;t) \\
D_2(M+1;t) & D_1(M;t) & \cdots & D_{-N+3}(M-N+2;t) \\
\vdots & \vdots & & \vdots  \\
D_N(M+N-1;t) & D_{N-1}(M+N-2;t) & \cdots & D_1(M;t)
\end{array}
\right|
\end{equation}
We can represent each element by a sum over the discrete time variable if 
$M\geq N$:
\begin{equation}
P(M,N,t)=p^N\left|
\begin{array}{cccc}
\sum_{t'=M-1}^{t-1}D_0(M-1;t') & \sum_{t'=M-2}^{t-1}D_{-1}(M-2;t') &
\cdots &  \sum_{t'=M-N}^{t-1}D_{1-N}(M-N;t')\\
\sum_{t'=M}^{t-1}D_1(M;t') & \sum_{t'=M-1}^{t-1}D_{0}(M-1;t') &
\cdots &  \sum_{t'=M-N+1}^{t-1}D_{2-N}(M-N+1;t')\\
\vdots & \vdots & & \vdots  \\
\sum_{t'=M+N-2}^{t-1}D_{N-1}(M+N-2;t') & \sum_{t'=M+N-3}^{t-1}D_{N-2}(M+N-3;t') 
& \cdots & \sum_{t'=M-1}^{t-1}D_0(M-1;t')
\end{array}
\right|
\end{equation}

Instead of the partial integration we perform partial summation here
and use the identity
\begin{equation}
\label{partialsum}
\sum_{t'=t_1}^{t-1} b_{t'} (a_{t'+1}-a_{t'}) = 
-\sum_{t'=t_1}^{t-1} a_{t'+1}(b_{t'+1}-b_{t'}) + a_{t}b_{t} - a_{t_1}b_{t_1}.
\end{equation}
for the second row with $a_t=t-1$:
\begin{align}
\sum_{t'=x}^{t-1}D_\nu (x,t') &= -p\sum_{t'=x}^{t-1}t'D_{\nu-1}(x-1,t') + 
(t-1)D_\nu(x,t) - (x-1)D_\nu(x,x) \cr
&= -p\sum_{t'=x-1}^{t-1}t'D_{\nu-1}(x-1,t') + (t-1)D_\nu(x,t).
\end{align}
For the third row we go further and apply (\ref{partialsum}) again
with $a_t=t(t-1)/2$:
\begin{multline}
-p\sum_{t'=x-1}^{t-1}t'D_{\nu-1}(x-1,t') + 
(t-1)D_\nu(x,t)= \cr
p^2\sum_{t'=x-1}^{t-1}\frac{t'(t'+1)}{2}D_{\nu-2}(x-2,t')
-p\frac{t(t-1)}{2}D_{\nu-1}(x-1,t)  + 
p\frac{(x-1)(x-2)}{2}D_{\nu-1}(x-1.x-1)+ (t-1)D_\nu(x,t)= \cr
p^2\sum_{t'=x-2}^{t-1}\frac{t'(t'+1)}{2}D_{\nu-2}(x-2,t')
-p\frac{t(t-1)}{2}D_{\nu-1}(x-1,t)  + (t-1)D_\nu(x,t).
\end{multline}
In the fourth row we perform the partial summation once more with
$a_t=(t-1)t(t+1)/3!$ and similarly for all the rows. Finally after
adding suitable linear combination of the first $n-1$ rows to the
$n$-th one we get
\begin{multline}
P(M,N,t)= \cr
\frac{p^\frac{N(N+1)}{2}(-1)^{\left[\frac{N}{2}\right]}}{0!1!\cdots (N-1)!}
\left|
\begin{array}{cccc}
\sum_{t'=M-1}^{t-1}D_0(M-1;t') & \sum_{t'=M-2}^{t-1}D_{-1}(M-2;t') &
\cdots &  \sum_{t'=M-N}^{t-1}D_{1-N}(M-N;t')\\
\sum_{t'=M-1}^{t-1}t'D_0(M-1;t') & \sum_{t'=M-2}^{t-1}t'D_{-1}(M-2;t') &
\cdots &  \sum_{t'=M-N}^{t-1}t'D_{1-N}(M-N;t')\\
\vdots & \vdots & & \vdots  \\
\sum_{t'=M-1}^{t-1}t'^{N-1}D_0(M-1;t') & \sum_{t'=M-2}^{t-1}t'^{N-1}D_{-1}
(M-2;t') & \cdots & \sum_{t'=M-N}^{t-1}t'^{N-1}D_{1-N}(M-N;t')
\end{array}
\right|
\end{multline}
We can set all the lower limits of the sums to 0 and write
\begin{equation}
P(M,N,t)= 
\frac{p^\frac{N(N+1)}{2}(-1)^{\left[\frac{N}{2}\right]}}{0!1!\cdots (N-1)!}
\sum_{t_1,t_2, \cdots ,t_N=0}^{t-1} t_1^0 t_2^1 t_3^2 \cdots t_N^{N-1}
\left|\begin{array}{cccc}
D_0(M-1;t_1) & D_{-1}(M-2;t_1) & \cdots &  D_{1-N}(M-N;t_1)\\
D_0(M-1;t_2) & D_{-1}(M-2;t_2) & \cdots &  D_{1-N}(M-N;t_2)\\
\vdots & \vdots & & \vdots  \\
D_0(M-1;t_N) & D_{-1}(M-2;t_N) & \cdots &  D_{1-N}(M-N;t_N)
\end{array}
\right|.
\end{equation}
Adding suitable linear combination of the first $n-1$ columns to the
$n$-th one we get (by using (\ref{Dqleq0}))
\begin{equation}
P(M,N,t)= 
\frac{p^\frac{N(N+1)}{2}(-1)^{\left[\frac{N}{2}\right]}}{0!1!\cdots (N-1)!}
\sum_{t_1,t_2, \cdots ,t_N=0}^{t-1} t_1^0 t_2^1 t_3^2 \cdots t_N^{N-1}
\left|\begin{array}{cccc}
D_0(M-1;t_1) & D_{0}(M-2;t_1) & \cdots &  D_{0}(M-N;t_1)\\
D_0(M-1;t_2) & D_{0}(M-2;t_2) & \cdots &  D_{0}(M-N;t_2)\\
\vdots & \vdots & & \vdots  \\
D_0(M-1;t_N) & D_{0}(M-2;t_N) & \cdots &  D_{0}(M-N;t_N)
\end{array}
\right|.
\end{equation}

Using the explicit form of $D_0$ (\ref{D0}) one arrives at
\begin{multline}
P(M,N,t)= 
\frac{p^{MN}(1-p)^{-MN+\frac{N(N+1)}{2}}(-1)^{\left[\frac{N}{2}\right]}}{0!1!
\cdots (N-1)!(M-1)!(M-2)!\cdots (M-N)!}\sum_{t_1,t_2, \cdots ,t_N=0}^{t-1} 
(1-p)^{\sum_{j=1}^N t_j}
t_1^0 t_2^1 t_3^2 \cdots t_N^{N-1} 
\times\cr
\left|\begin{array}{cccc}
t_1(t_1-1)\cdots (t_1-M+2) & t_1(t_1-1)\cdots (t_1-M+3) & \cdots & t_1(t_1-1)
\cdots (t_1-M+N+1) \\
t_2(t_2-1)\cdots (t_2-M+2) & t_2(t_2-1)\cdots (t_2-M+3) & \cdots & t_2(t_2-1)
\cdots (t_2-M+N+1) \\
\vdots & \vdots & & \vdots  \\
t_N(t_N-1)\cdots (t_N-M+2) & t_N(t_N-1)\cdots (t_N-M+3) & \cdots & t_N(t_N-1)
\cdots (t_N-M+N+1) 
\end{array}
\right| =\cr \cr
\frac{p^{MN}(1-p)^{-MN+\frac{N(N+1)}{2}}(-1)^{\left[\frac{N}{2}\right]}}
{\prod_{j=1}^N (j-1)!(M-j)!} \times \cr
\sum_{t_1,t_2, \cdots ,t_N=0}^{t-1} 
(1-p)^{\sum_{j=1}^N t_j}
t_1^0 t_2^1 t_3^2 \cdots t_N^{N-1} 
\prod_{j=1}^N t_j(t_j-1)(t_j-2)\cdots (t_j-M+N+1)
\left|\begin{array}{cccc}
t_1^{N-1}  & t_1^{N-2} & \cdots & 1 \\
t_2^{N-1}  & t_2^{N-2} & \cdots & 1 \\
\vdots & \vdots & & \vdots  \\
t_N^{N-1}  & t_N^{N-2}  & \cdots & 1 
\end{array}
\right| =\cr \cr
\frac{p^{MN}(1-p)^{-MN+\frac{N(N+1)}{2}}}{\prod_{j=1}^N j!(M-j)!}
\sum_{t_1,t_2, \cdots ,t_N=0}^{t-1} 
\prod_{j=1}^N t_j(t_j-1)(t_j-2)\cdots (t_j-M+N+1)(1-p)^{t_j}
\left|\begin{array}{cccc}
t_1^{N-1}  & t_1^{N-2} & \cdots & 1 \\
t_2^{N-1}  & t_2^{N-2} & \cdots & 1 \\
\vdots & \vdots & & \vdots  \\
t_N^{N-1}  & t_N^{N-2}  & \cdots & 1 
\end{array}
\right|^2.
\end{multline}
Finally we get
\begin{equation}
\label{Presult}
P(M,N,t)= \frac{p^{MN}(1-p)^{-MN+\frac{N(N+1)}{2}}}{\prod_{j=1}^N
  j!(M-j)!} \sum_{t_1,t_2, \cdots ,t_N=0}^{t-1} \left(\prod_{j=1}^N
  \left(\prod_{k=0}^{M-N-1}(t_j-k)\right)(1-p)^{t_j}\right)
\prod_{i<j}(t_i-t_j)^2.
\end{equation}
One can see that in the limit $t=\tilde t/p, p\to 0, \tilde
t=\text{const.}$ (\ref{Presult}) is equivalent to (\ref{result}) as
expected. 

\begin{figure}
\label{fig:path}
\epsfig{file=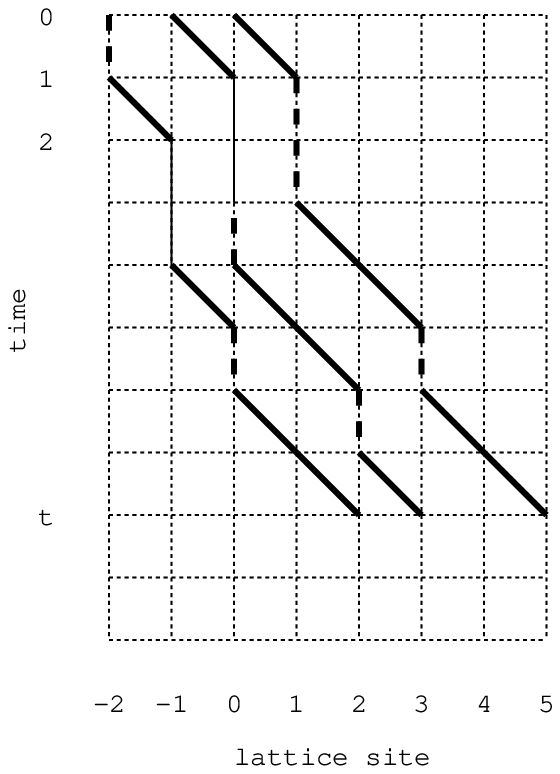,width=6truecm} \quad
\epsfig{file=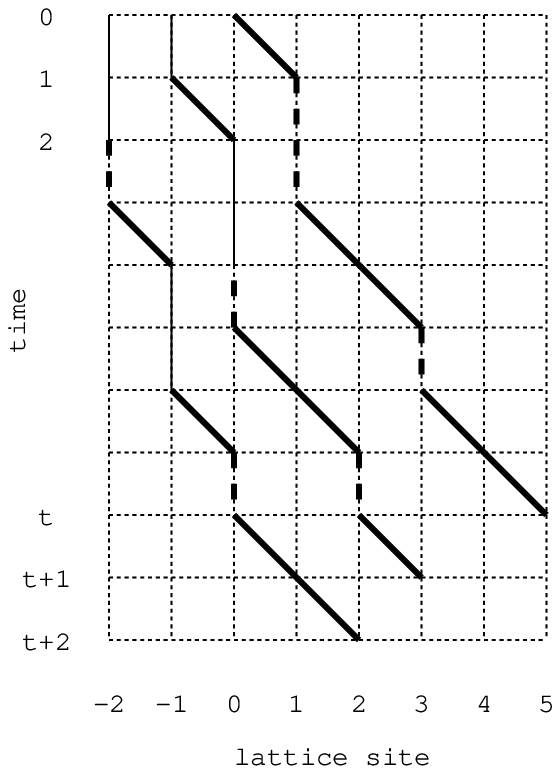,width=6truecm}
\caption{The first figure shows a possible history of three particles from 
time 0 to time $t$. We adjust probability $p$ to the diagonal lines 
(thick solid lines) which correspond to particle hopping. The weight of the 
thick dashed vertical lines is $1-p$ while the thin solid vertical lines have 
weight 1. The second figure shows the corresponding path configuration of the 
DTASEP. The path of the second (third \dots) particle is shifted by one 
(two \dots) time steps.}
\end{figure}

Expression (\ref{Presult}) can be written as 
\begin{equation}
\label{Presult2}
  P(M,N,t)= Z(M,N)^{-1} \sum_{t_1,t_2, \cdots ,t_N=0}^{t-1-M+N}
  \prod_{j=1}^N \left(\binom{t_j+M-N}{M-N}(1-p)^{t_j}\right)
  \prod_{i<j}(t_i-t_j)^2.
\end{equation}
In the latter form $P(M,N,t)$ is similar to the same quantity of the
usual discrete time ASEP (DTASEP) (see Proposition 1.3 of \cite{Joha00}).
Namely:
\begin{equation}
\label{rel}
  P(M,N,t)=P_\text{DTASEP}(M,N,t+N-1)
\end{equation}
This correspondence can be seen in a graphical representation of the 
dynamics (Fig. 1). The statistical weight
coming from the path of the first particle is the same in both
processes since its dynamics is identical in the two cases. The weight
of the path of the $N$th particle from $0$ to $t$ is the same as that
of the normal DTASEP from $0$ to $t+N-1$ which explains the relation
(\ref{rel}) (see also figure \ref{fig:path}). However, this
correspondence is valid only in the case of this special initial
condition.

From the geometric interpretation it follows that (\ref{rel}) is
valid for all values of $M$ and $N$, although we derived this (on the
level of formulas) only for $M\geq N$ (note that Proposition 1.3 of
\cite{Joha00} as well as (\ref{Presult2}) is valid only for $M\geq
N$). The DTASEP has particle-hole symmetry which implies
\begin{equation}
\label{symm1}
P_\text{DTASEP}(M,N,t)=P_\text{DTASEP}(N,M,t),
\end{equation} 
from this we get
\begin{equation}
\label{symm2}
  P(M,N,t)=P_\text{DTASEP}(M,N,t+N-1)=P_\text{DTASEP}(N,M,t+N-1)=P(N,M,t+N-M).
\end{equation}

\subsection{Asymptotic form of the distribution function}

Knowing the derivation of the asymptotic form of
$P_\text{DTASEP}(M,N,t)$ (see section 3 of \cite{Joha00}) it is
rather easy to obtain similar results for this process. We repeat the
result of Johansson for the discrete-time TASEP with parallel update
with our notations: 
\begin{equation}
\label{asym}
\lim_{N\to\infty}P_\text{DTASEP}\left([\gamma N],N,N\omega_\text{DTASEP}(\gamma,p)+N^{1/3}
\sigma(\gamma,p)s\right)=
F_\text{GUE}(s)
\end{equation}
with 
\begin{gather}
  \omega_\text{DTASEP}(\gamma,p)=
  \frac{\left(1+\sqrt{(1-p)\gamma}\right)^2}{p}+\gamma, \\
  \sigma(\gamma,p)=\frac{(1-p)^{1/6}\gamma^{-1/6}}{p}
  \left(\sqrt{\gamma}+\sqrt{1-p}\right)^{2/3}
  \left(1+\sqrt{(1-p)\gamma}\right)^{2/3}.
\end{gather}
This formula is derived for $\gamma\geq 1$ in
\cite{Joha00} but it is easy to show that it is valid also for
$0<\gamma<1$. Using (\ref{symm1}) one gets 
\begin{equation}
  \lim_{N\to\infty}P\left(\gamma N,N,N\omega_\text{DTASEP}(1/\gamma,p)\gamma + 
 N^{1/3}\sigma(1/\gamma,p)\gamma^{1/3}s\right)=
  F_\text{GUE}(s),
\end{equation}
which is identical to (\ref{asym}) since
$\omega_\text{DTASEP}(1/\gamma,p)\gamma=\omega_\text{DTASEP}(\gamma,p)$
and $\sigma(1/\gamma,p)\gamma^{1/3}=\sigma(\gamma,p)$.

To obtain the corresponding asymptotic result for the
fragmentation process one goes through the same steps 
as Ref. \cite{Joha00} (or simply use (\ref{rel})) and (\ref{asym}). One finds
\begin{equation}
  \lim_{N\to\infty}P\left([\gamma N],N,N\omega(\gamma,p) +
    N^{1/3}\sigma(\gamma,p)s\right)= F_\text{GUE}(s)
\end{equation}
with
\begin{equation}
  \omega(\gamma,p)=
  \frac{\left(1+\sqrt{(1-p)\gamma}\right)^2}{p}+\gamma-1=
  \frac{\left(\sqrt{1-p}+\sqrt{\gamma}\right)^2}{p}.
\end{equation}
This result is valid 
for any $\gamma >0 $. The corresponding identity is
$\omega(1/\gamma,p)\gamma-1+\gamma=\omega(\gamma,p)$.
We remark that amplitude $\sigma$ of the deviation is the
same for both models.

\section{Conclusions}

We have shown that the distribution of the time-integrated current for
the TASEP as well as for the totally asymmetric fragmentation process
can be obtained using Bethe ansatz. To show this one uses a
determinant representation of the Bethe wave function which solves the
master equation for a finite number of particles. After appropriate
scaling the current distribution is given by the distribution of the
largest eigenvalue of a random matrix ensemble.  This observation may
lead to better understanding of the relation between the random matrix
theory and the Bethe ansatz and suggests that also other integrable
hopping processes may be treated in a similar fashion. The main task
that remains is the derivation of suitable determinant representations
for conditional probabilities for such processes.  In the scaling
limit of the fragmentation process the distribution of the
time-integrated current around its mean converges to the same
Tracy-Widom distribution for the Gaussian unitary ensemble found
previously for the TASEP, thus confirming universality of this
quantity.

\begin{acknowledgments}
G.M.S. would like to thank H. Spohn for useful discussions. A.R. acknowledges
financial support by the Deutsche Forschungsgemeinschaft under grant number
Schu827/6-1.
\end{acknowledgments}

\vspace{1cm}
{\bf Note added:} After completion of this work we learned that the result
of Sec. (\ref{Calculation}) has been found also by Nagao and Sasamoto,
cond-mat/0405321.

\appendix
\section{$F$ functions}
\label{F_functions}

Definition:
\begin{align}
F_p(n;t)&=\frac{1}{2\pi} \int_{-\pi}^{\pi} e^{-(1-e^{-ik})t} 
\left(1-e^{i(k+i0)}\right)^{-p} e^{ikn} dk \\
&=\frac{1}{2\pi i} \oint_{|z|=1-0} e^{-(1-z^{-1})t} (1-z)^{-p} z^{n-1} dz.
\end{align}
The integral is taken in a way that the pole at $z=1$ should not be
taken into account.

Differentiation, integration: 
\begin{gather}
\frac{d}{dt}F_p(n;t)=F_{p-1}(n-1;t) \\
\label{int}
\int_{t_1}^{t_2}F_p(n;t)=F_{p+1}(n+1,t_2)-F_{p+1}(n+1,t_1)
\end{gather}

Summation:
\begin{gather}
\label{sum1}
\sum_{n=n_1}^{n_2} F_p(n;t) = F_{p+1}(n_1;t)-F_{p+1}(n_2+1;t) \\
\label{sum2}
F_p(n;t) = F_{p+1}(n;t) - F_{p+1}(n+1;t) \\
\label{sum3}
\sum_{n=n_1}^{\infty} F_p(n;t)= F_{p+1}(n_1;t)
\end{gather}

For $p\leq 0$ and $n\geq 0$:
\begin{equation}
\label{pneg}
F_p(n;t) = e^{-t} \sum_{m=0}^{-p} (-1)^m \binom{-p}{m}
\frac{t^{m+n}}{(m+n)!}
\end{equation}
$F_p(n;t)$ can be written as a (finite or infinite) sum also in other
regions of the $(n,p)$ parameter space (see \cite{Schu97}) but those
formulas are not used here.  

For $n\geq 0$ (and any $p$):
\begin{equation}
F_p(n;0) = \delta_{n,0}
\end{equation}

\section{$D$ functions}
\label{appendix:D}

Definition:
\begin{align}
D_q(n,t) &= \frac{1}{2\pi}\int_0^{2\pi} dk \left(1-p+pe^{-ik}\right)^t 
\left(1-e^{i(k+i0)}\right)^{-p} e^{ikn} \\
& = \frac{1}{2\pi i} \oint_{|z|=1-0} dz 
\left( 1-p+\frac{p}{z}\right)^t (1-z)^{-q} z^{n-1}
\end{align}

Discrete time derivative and summation for $t$:
\begin{gather}
D_q(n,t+1)-D_q(n,t) = pD_{q-1}(n-1,t) \\
\sum_{t=t_1}^{t_2}D_q(n,t) =\frac{1}{p}\left(D_{q+1}(n+1,t_2+1)-D_{q+1}
(n+1,t_1)\right)
\end{gather}

Summation for $n$:
\begin{gather}
\label{Dsum1}
\sum_{n=n_1}^{n_2} D_q(n;t) = D_{q+1}(n_1;t)-D_{q+1}(n_2+1;t) \\
\label{Dsum2}
D_q(n;t) = D_{q+1}(n;t) - D_{q+1}(n+1;t) \\
\label{Dsum3}
\sum_{n=n_1}^{\infty} D_q(n;t)= D_{q+1}(n_1;t)
\end{gather}

Explicit form of $D_q(n,t)$ for $n,t\geq 0$:
\begin{equation}
\label{D0}
D_0(n,t)= \begin{cases}0 & t<n \cr 
\binom{t}{n}(1-p)^{t-n}p^n & 0\leq n \leq t \cr 
0 & n \leq 0 \end{cases}
\end{equation}
\begin{gather}
D_{q>0}(n,t)=\sum_{j=0}^\infty \binom{q+j-1}{j}D_0(n+j,t) \\
\label{Dqleq0}
D_{q\leq 0}(n,t)=\sum_{j=0}^{-q} \binom{-q}{j}(-1)^j D_0(n+j,t) \\
D_q(n,n)=D_0(n,n)=p^n
\end{gather}

\end{document}